\documentclass[conference,10pt]{IEEEtran}
\usepackage{cite}     
\usepackage{spconf}
\usepackage{graphicx}  
\usepackage{url}
\usepackage{times}
\usepackage{amsmath}   
\usepackage{pdflscape}
\usepackage{balance}
\usepackage{url}
\usepackage{amsfonts,amsthm,amssymb}
\usepackage{mathtools}
\usepackage{nicefrac}
\usepackage[none]{hyphenat}
\usepackage{array}
\usepackage{fontenc}
\usepackage{color}
\usepackage{etoolbox}
\usepackage{bm}
\usepackage{comment}
\usepackage{amsmath,amsthm,amssymb, amsfonts}
\usepackage{amssymb,mathtools}
\usepackage[normalem]{ulem}
\usepackage{nomencl}
\usepackage{accents}
\usepackage{algorithm, algorithmic}

\newcommand{\argmax}{\arg\!\max}

\title{IEEE 802.11ad-Aided 5-D Sensing with a UAV Swarm in Urban Environment}

\name{Akanksha~Sneh\textsuperscript{1},   Shobha~Sundar~Ram\textsuperscript{1} and Kumar~Vijay~Mishra\textsuperscript{2}
 }
\address{\textsuperscript{1}Indraprastha Institute of Information Technology Delhi, New Delhi 110020 India\\
\textsuperscript{2}United States DEVCOM Army Research Laboratory, Adelphi, MD 20783 USA}

\begin{document}
\ninept

\maketitle

\begin{abstract} 
Aerial base stations mounted on unmanned aerial vehicles (UAVs) support next-generation wireless networks in challenging environments such as urban areas, disaster zones, and remote locations. Further, UAV swarms overcome the challenges of limited battery life and other operational constraints of a single UAV. However, tracking mobile users on the ground by each UAV and the corresponding synchronization between the UAVs is a significant issue that must be addressed before this framework can be deployed in reality. Incorporating additional sensing capabilities to facilitate this additional requirement would introduce significant overhead in terms of hardware, cost, and power to each UAV. Instead, we propose an integrated sensing and communications-enabled swarm UAV system, based on the millimeter-wave IEEE 802.11ad protocol. Further, we show that our proposed system is capable of five-dimensional (5-D) ground target sensing (range, Doppler velocity, azimuth, elevation, and polarization) in an urban environment. Numerical experiments using realistic models demonstrate and validate the performance of 5-D sensing using our proposed 802-11ad-aided UAV system.
\end{abstract}
%
\section{Introduction}
Recently, there has been increasing interest in deploying aerial base stations mounted on aircraft or unmanned aerial vehicles (UAVs) to enhance next-generation wireless networks \cite{zheng2016Wir,  mozaffari2016efficient}. These aerial platforms offer several advantages, including rapid deployment, flexible repositioning, and access to line-of-sight (LOS) propagation links. These features make them particularly valuable in challenging scenarios, such as densely populated urban areas \cite{zema2017unmanned}, natural disaster zones, or locations with physical obstructions like mountains, forests, or glaciers \cite{rong2021noncontact,valavanis2015handbook}, where traditional ground-based infrastructure struggles to provide reliable connectivity. 

However, one of the primary challenges in the existing UAV infrastructure is the limited durability due to constrained battery power and operational constraints \cite{challita2019interference}. 
To address this issue, researchers have focused on several strategies, such as developing low-power hardware \cite{xiao2021machine},  coordinating swarms of multiple drones to work as multiple subswarms for several operations \cite{ren2022swarm}, and optimizing path planning of the flights \cite{wang2019autonomous}. Thus, the efficient use of resources is critical in these systems, where the limited battery life and minimal hardware necessitate careful power management and operational efficiency, often requiring advanced control algorithms and innovative energy-saving techniques. 

Further, radars and wireless communication systems on UAVs are increasingly utilized for tasks like sensing targets that may be obscured as well as serving as communication infrastructure in inaccessible areas \cite{wu2021comprehensive}. There is a growing trend toward integrating sensing and communication (ISAC) functionalities into common hardware, spectrum, and waveform to conserve power, reduce physical space, save spectrum, and lower costs \cite{alaee2019discrete, dokhanchi2019mmwave, mishra2019toward, duggal2019micro}. This integrated approach not only simplifies the design but also enhances the versatility of UAVs, enabling them to perform complex missions with minimal resources. This has led to significant interest in the development of transceivers that can simultaneously handle radar and communication tasks using a shared spectrum, allowing UAVs to perform surveillance and communication roles concurrently \cite{mishra2019toward}. 

Further, there is a requirement for high data rates with large bandwidth for next-generation communication. The millimeter-wave (mmW) band has appeared as a viable solution for UAVs as it allows for compact airborne hardware and provides extensive bandwidth, enabling high data rates with finer radar range resolution. These mmW-equipped UAV systems are envisioned to seamlessly connect with dynamic ground users, such as pedestrians and vehicles, offering a versatile and powerful tool for next-generation wireless communication networks. As these technologies evolve, aerial base stations are expected to bridge connectivity gaps, enabling intelligent transportation systems in smart cities and marking a significant step forward in the evolution of wireless communication infrastructure.

Over the recent years, ISAC systems for mmW/Terahertz spectrum have been extensively explored \cite{vargas2023dual,nayemuzzaman2024co,elbir2024spatial,jacome2023multi,wei2024ris,elbir2021terahertz}.
Lately, for ISAC systems, existing communication protocols for ISAC systems have been exploited for the sensing functionality \cite{ mishra2019toward,kumari2018ieee,kumari2015investigating,chen2024integrated}. Particularly, IEEE 802.11ad protocol, which operates at a carrier frequency of 60GHz, is a promising candidate for inter-vehicular communication links \cite{duggal2019micro,duggal2020doppler,sneh2023ieee,sneh2024beam}. The Golay complementary sequences in the channel estimation field of the preamble of the IEEE 802.11ad frame possess perfect auto-correlation properties that can be utilized for the range estimation of the targets. Moreover, unlike traditional omnidirectional beam patterns, IEEE 802.11ad employs digital beamforming, which directs narrow beams toward specific mobile users. In \cite{ram2022uav},  the IEEE 802.11ad-based ISAC UAV was investigated for the radar sensing of pedestrians and vehicles on the road. 

ISAC-enabled UAV swarm has been investigated in \cite{zhou2022integrated} for multiple targets tracking through cyber-twin-based AI technology. ISAC-based swarm UAV systems offer significant advantages over single UAV systems, including enhanced coverage, scalability, and azimuth-elevation information through coordinated multi-angle data collection. It provides increased reliability, allowing the swarm to maintain operations even if individual UAVs fail. All of the works based on ISAC-enabled UAV swarm present in the literature have been explored for sub-6 GHz frequency. 
In this paper, we propose an IEEE 802.11ad-based ISAC-enabled UAV swarm hovering at one place for an mmW urban environment as shown in Fig.~\ref{fig:sysmod} for ground target five-dimensional  (5-D) sensing that includes range, Doppler velocity, azimuth, elevation, and polarization.

The rest of the paper is organized as follows. In the next section, we present the signal model 
followed by 5-D sensing algorithms in  Section~\ref{sec:5d}. We validate our model and methods through a realistic simulation setup and numerical experiments in Section~\ref{sec:Results}. We conclude in Section~\ref{sec:Conclusion}. 

Throughout this paper, scalar variables, vectors, and matrices are represented with regular, lower-case bold, and upper-case bold characters, respectively. Symbols $\otimes$, $\odot$, and $^\mathbb{H}$ denote the convolution, dot product, and complex conjugate transpose operations, respectively. 

\section{Signal Model}
\label{sec:signalmodel}
We introduce the signal model of the IEEE 802.11ad-based ISAC-enabled UAV swarm transceiver. Further, we present the radar signal processing in order to perform the 5-D sensing of the potential targets.

\begin{figure}[t]
    \centering
    \includegraphics[width=1.0\columnwidth]{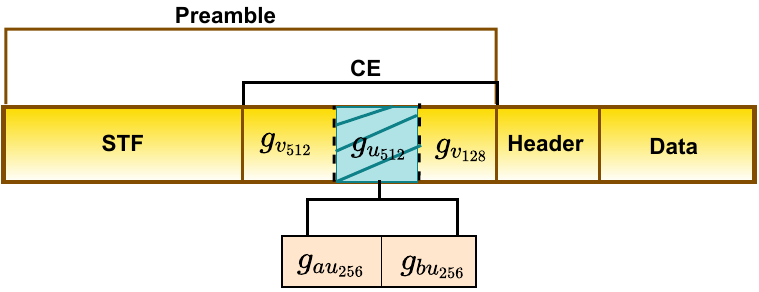}
    \caption{{\color{black}IEEE 802.11ad PHY frame with radar waveform embedded in it.}}
    \label{fig:phy_packet}   
\end{figure}

\textbf{{Transmitted signal}}: The IEEE 802.11ad PHY frame structure comprises a preamble, header, and data, as shown in Fig.~\ref{fig:phy_packet}.
The 512-Golay sequence $\mathbf{g}_{u_{512}}$ (shown in pink color) in the channel estimation (CE) of the preamble of the PHY packet forms the radar waveform. The $P$ discrete samples of the radar waveform $y[n]$ within one pulse repetition interval ($T_{P}$) corresponding to each $q^{th}$ packet  are converted to analog signals by passing them through digital-analog converter as the Dirac-delta function $\delta(\cdot)$:
\par\noindent
\begin{align}
    \mathbf{y}(t) =\sum_{q=0}^{Q-1}\sum_{p=0}^{P-1}y[pT_s]\delta\left(t- pT_s - qT_{P}\right),
\end{align}\normalsize
where $T_s$ corresponds to the sampling time. These analog signals are amplified with $\sqrt{E}_s$ energy per sample, passed through a transmit-shaping filter $l(t)$, and then upconverted to an mmW frequency $f_o$. We assume that the $N$ UAVs correspond to the $N$ antenna elements arranged in a circular array (UCA). The upconverted signals are passed through analog beamforming through UCA:
\par\noindent
\begin{align}
\label{eq:tx_weights}
    \mathbf{Y}_{{tx}}(t) = \mathbf{w_{UCA}}(\sqrt{E}_s(\mathbf{y}(t) \otimes l(t) )).
\end{align}\normalsize
Here, the $\mathbf{w_{UCA}}$ weight vector $\mathbf{w} \in \mathbb{C}^{N \times 1}$  is assigned to antenna elements/ UAVs according to the quasi-omni direction, as the targets have not been localized initially. In order to sense the polarization at the receiver with respect to the target, we transmit horizontally and vertically polarized electric fields:
\par\noindent
\begin{align}
\label{eq:tx_polarized}
    \mathbf{S}_{{tx_{H}}}(t) =  \xi_H\mathbf{Y}_{{tx}}(t),\\
     \mathbf{S}_{{tx_{V}}}(t) =  \xi_V\mathbf{Y}_{{tx}}(t),
\end{align}\normalsize
where $\xi_H$ and $\xi_V$ denote the horizontal and vertical element pattern of each antenna, respectively. \\
\begin{figure}[t]
    \centering
    \includegraphics[width=1.0\columnwidth]{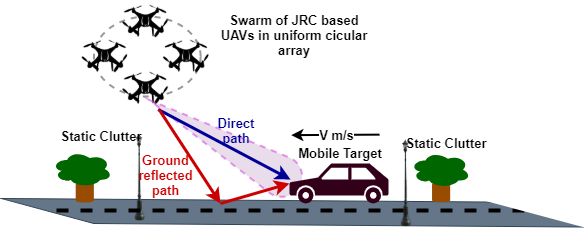}
    \caption{ System model showing a swarm of UAVs sending a directional beam to the mobile target.}
    \label{fig:sysmod}
\end{figure}
\textbf{{Received signal}}: We assume M targets present in the environment. Then, the incident horizontal and vertical electric fields falling on the targets is
 \par\noindent
\begin{align}
    \mathbf{S}_{{inc_H}}(t) = \sum_{m=1}^{M}\mathbf{U}_{tx_{m_{H}}} \odot \mathbf{S}_{{tx_{H}}}(t), \\
     \mathbf{S}_{{inc_V}}(t) = \sum_{m=1}^{M}\mathbf{U}_{tx_{m_{V}}} \odot \mathbf{S}_{{tx_{V}}}(t),
\end{align}\normalsize \\ 
Note that since the UAVs are at a certain height with respect to the ground, we assume that the transmitting signal falling on the target follows two paths: one is the direct LOS path from the transmitter to the targets, and the other is the ground reflected path as shown in Fig.~\ref{fig:sysmod}. Thus, the propagation vector $\mathbf{u}_{tx_{m,n}}$ with respect to each $n^{th}$ antenna element and $m^{th}$ target incorporates the direct and ground-reflect paths and can be expressed  for horizontal and vertical polarizations (H/V) as 
 \par\noindent
\begin{align}
    \mathbf{u}_{tx_{(m,n)_{H/V}}}(t) = \frac{1}{\sqrt{4\pi} {r_{(m,n)}(t)}}e^{-jk{r_{(m,n)}(t)}} + \\ \frac{\Gamma_{H/V} }{\sqrt{4\pi} {r'_{(m,n)}(t)}}e^{-jk{r'_{(m,n)}(t)}}.
\end{align}\normalsize \\
Here, $r_{(m,n)}$ and $r'_{(m,n)}$ represent the direct LOS distance and ground reflect distance of $m^{th}$ target to  $n^{th}$ antenna element respectively and $k = \frac{2\pi}{\lambda}$ is the propagation constant. $\Gamma_{H/V}$ denotes the reflection coefficient for H/V polarization. Finally, the incident electric fields are scattered from the targets and are collected at the receiver antenna array:
      \begin{align}
      \label{eq:rx_radar}
      \begin{bmatrix}
         {{\mathbf{s}}}_{rx_{(m,n)_{H}}(t)} \\ 
        {{\mathbf{s}}}_{rx_{(m,n)_{V}}(t)}  \\
         \end{bmatrix}
   = \sum_{m=1}^M \begin{bmatrix}
  \eta_{{(m,n)}_H}(t) \\ \eta_{{(m,n)}_V}(t)
   \end{bmatrix}
   \begin{bmatrix}
  \sigma_{m_{HH}} & \sigma_{m_{HV}}\\
    \sigma_{m_{VH}}  & \sigma_{m_{VV}}  \\
  \end{bmatrix} \\
   \begin{bmatrix}
      \mathbf{u}_{rx_{(m,n)_{H}}(t)}\ \odot \mathbf{s}_{{inc_{(m,n)_{H}}}}(t) \\ 
        \mathbf{u}_{rx_{(m,n)_{V}}(t)}\odot \mathbf{s}_{{inc_{(m,n)_{V}}}}(t)\\
         \end{bmatrix} 
\mathbf{\zeta}(t) +\mathbf{\beta}(t) ,
\end{align} \normalsize
where, $ \mathbf{U}_{rx_m(t)}$ comprise the propagation vectors with respect to the receiver. The received electric fields are then multiplied with the two-way propagation path loss factor $\eta$ and scattering matrix comprising four different radar cross-section (RCS) components of the environment comprising horizontal-horizontal (H-H) polarization component $\sigma_{HH}$, vertical-vertical (V-V)  polarization component $\sigma_{VV}$ and horizontal-vertical cross-polarization (H-V and V-H) components, $\sigma_{HV}$ and $\sigma_{VH}$. $\mathbf{\zeta}$ and $\mathbf{\beta}$ denote ground clutter present in the environment and the complex additive white Gaussian noise at the receiver, respectively. We can obtain the received signal matrix $ {{\mathbf{Y}}}_{{rx}_{H/V}}$ from the electric field:
   \begin{align}
    \label{eq:rx_radar2}
      \begin{bmatrix}
          {{\mathbf{Y}}}_{{rx}_{H}}(t) \\ 
        {{\mathbf{Y}}}_{{rx}_{V}}(t)  \\
         \end{bmatrix} = \left(\frac{\lambda}{\sqrt{4\pi}}\right) \begin{bmatrix}
           {{\mathbf{S}}}_{{rx}_{H}}(t) \\ 
       {{\mathbf{S}}}_{{rx}_{V}}(t) \\
         \end{bmatrix}.
   \end{align} \normalsize

\begin{algorithm}[htbp] 
	\caption{5-D localization of multiple targets}
	\label{alg:5d_sensing}
	\begin{algorithmic}[1]
		\STATE \textbf{Input:} $N$, $K$, ${{\mathbf{Y}}}_{{rx}_{H/V}}$
		\STATE \textbf{Output:} $r_{k_{H/V}}$, $f_{k_{H/V}}$, $\phi_{k_{H/V}}$, $\theta_{k_{H/V}}$; $\forall  1 \leq k \leq K$  
		\FOR{$n =1, 2 \ldots N, $}
   
         \STATE  ${{\mathbf{Y}}}_{{rx}_{n_{H/V}}}[p_{\tilde{f}},q_{\tilde{f}}]$: Perform FFT on  ${{\mathbf{Y}}}_{{rx}_{n_{H/V}}}[p,q]$.
         \STATE $\mathbf{X}_{n_{{H/V}}}[r_{\tilde{f}},f]$: Perform matched filtering across fast time samples and 1D FFT across slow time samples
         \STATE $\mathbf{X}_{n_{{H/V}}}[r,f]$: Perform IFFT on $\mathbf{X}_{n_{{H/V}}}[r_{\tilde{f}},f]$ to obtain 2D range Doppler ambiguity function.
         \STATE $ [r_{m^*},f_{m^*}]_{n_{H/V}}$: Obtain the range and Doppler bin corresponding to the highest strength target.
         \FOR {$k =1, 2 \ldots K, $}
         \STATE $[a_k, r_k, f_k]_{n_{H/V}}$: Perform CLEAN and obtain the highest peak in the 2D range-Doppler ambiguity map of  $k^{th}$ CLEAN iteration.
         \ENDFOR
         \ENDFOR
         \STATE $[\phi_k, \theta_k]_{H/V}$: Perform 2D MUSIC for azimuth and elevation estimation of all K targets. \hspace{2cm} $\vartriangleright$ \textbf{Algorithm 2}

 	\end{algorithmic}
\end{algorithm}


\begin{algorithm}[htbp] 
	\caption{Joint azimuth-elevation estimation through UCA of UAV swarm}
	\label{alg:2D_MUSIC}
	\begin{algorithmic}[1]
		\STATE \textbf{Input:} $N$, $K$, $\mathbf{X}_{n_{{H/V}}}[r_k,f_k]$
		\STATE \textbf{Output:} $\phi_{k_{H/V}}$, $\theta_{k_{H/V}}$ ; $\forall  1 \leq k \leq K$

  \FOR{$k =1, 2 \ldots K, $}

            \STATE Vectorize the data across  $N$ antenna elements for H/V polarization as shown: \\ $ \mathbf{x}_{k_{H/V}} = \left[\mathbf{X}_{1_{H/V}}[{r}_{k},f_k] \cdots,\mathbf{X}_{N_{H/V}}[{r}_{k},f_k]\right]$
           
		    \STATE $\mathbb{B}_{k_{H/V}}$: Obtain the covariance matrix of $\mathbf{x}_{k_{H/V}}$.
		    \STATE Perform eigenvector decomposition on $\mathbb{B}_{k_{H/V}}$ by QR-decomposition method.
		\STATE $\mathbb{E}_{k_{H/V}}$: Estimate the  the noise subspace.
              \FOR{$i =-180^o \ldots 180^o, $}
			\FOR{$j = 90^o \ldots 180^o, $}
            \STATE Obtain the array factor of UCA as shown in the below equation: 
            $\alpha_{k_{H/V}}[n] = e^{j k sin(i) cos(j-\phi_n)}$; $\phi_n$ is elevation of $n^{th}$ UAV.
            \STATE Generate the MUSIC spectrum as shown in the below equation: \\
            $\mathbb{P}_{k_{H/V}}[i,j] = \frac{1}{\alpha^\mathbb{H}_{k_{H/V}} \mathbb{E}_{k_{H/V}} \mathbb{E}^\mathbb{H}_{k_{H/V}} \alpha_{k_{H/V}}}$.
            \STATE $[\phi_k, \theta_k]_{H/V} = \argmax \mathbb{P}_{k_{H/V}}$
               \ENDFOR
            \ENDFOR
        \ENDFOR
	\end{algorithmic}
\end{algorithm}

 \begin{figure*}[ht]
    \centering
    \includegraphics[width=1\textwidth]{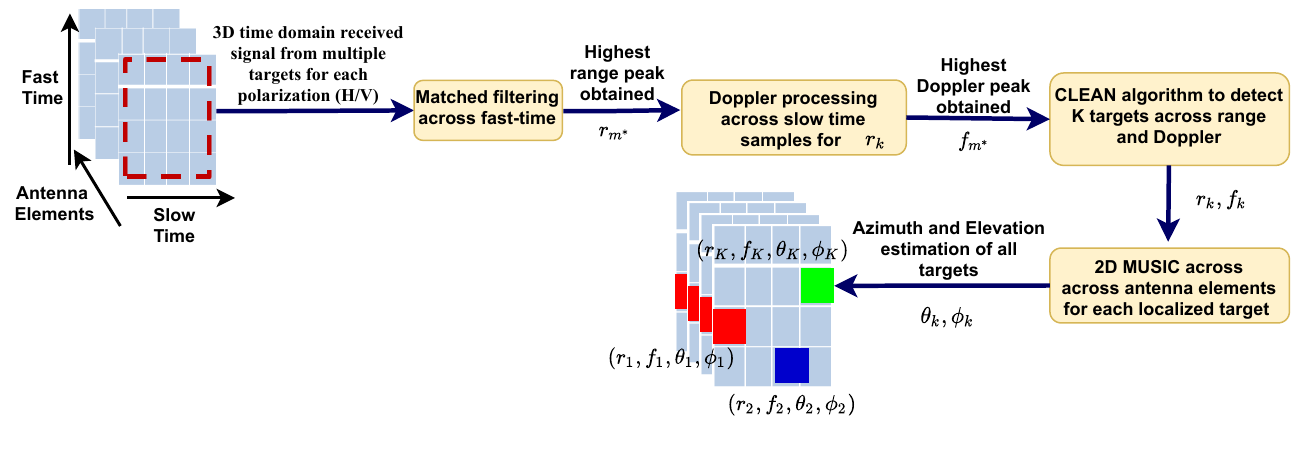}
    \caption{Block diagram of 5-D sensing: Input radar data cube for each polarization is first processed to obtain range-Doppler ambiguity plots. Multiple targets are detected using CLEAN. Corresponding azimuth and elevation angles of each target are estimated through 2D MUSIC.} 
    \label{fig:rsp_block}   
\end{figure*}

\section{5-D Sensing}
\label{sec:5d}
The received signals ${{\mathbf{Y}}}_{{rx}_{H/V}}(t)$ across the UCA of UAVs are collected and passed through the downconverter and analog-to-digital converter to convert them back to the baseband. Now, these downconverted and digitized signals are arranged in the form of a radar data cube of dimension $N \times P \times Q $  for each polarization that goes through a radar signal processing block as shown in Fig.~\ref{fig:rsp_block} for 5-D sensing of the targets. Firstly, matched filtering is carried out in the frequency domain across the fast-time samples. This is achieved by multiplying the samples with the complex conjugate of the original Golay sequences for each radar data cube as shown in Algorithm~\ref{alg:5d_sensing} [Line 1-5]. Subsequently, one-dimensional (1D) fast Fourier transform (FFT)is carried out for Doppler processing across the packets corresponding to slow-time samples. Next, inverse FFT is performed for each Doppler packet to convert back to them back to the time/range domain to obtain 2D range-Doppler ambiguity function $X_{n_{H/V}}[r,f]$ for each antenna element and polarization given in Algorithm~\ref{alg:5d_sensing} [Line 6]. Further,  the highest peak $r_{m^*}$ corresponding to the range sample $p_{m^*}$ of the highest strength target and the Doppler peak $f_{m^*}$ across the slow-time sample $q_{m^*}$ of the highest strength target are obtained as shown in Algorithm~\ref{alg:5d_sensing} [Line 7]. Note that the Doppler velocity of the peak for $f_{m^*}$ can be obtained as $v_{m^*} = \frac{f_{m^*} \lambda}{2}$, where $\lambda$ is the wavelength corresponding to $f_o$. Then, we localize multiple targets present in the environment through the CLEAN algorithm \cite{tsao1988reduction,ram2008through}. It is an iterative process wherein, at each step, the peak associated with the range and Doppler of a target is identified by subtracting the point spread response of the strongest target detected in the preceding iteration. The output of the CLEAN algorithm consists of all potential targets, $k=1:K$, obtained at each $k^{th}$ iteration of the CLEAN algorithm and their corresponding amplitudes and range-Doppler positions ($a_k, r_k,f_k$) as shown in Algorithm~\ref{alg:5d_sensing} [Line 8-11]. Note that some of the targets may not be mobile targets and may be static ground clutter in the environment, resulting in zero Doppler. Thus, we discard the zero Doppler peaks. Finally, we perform a 2D multiple signal classification (MUSIC) algorithm as shown in Algorithm~\ref{alg:2D_MUSIC}. Here, 2D MUSIC is performed on the indices of the 2D range-Doppler ambiguity map corresponding to the targets across all the antenna elements, and azimuth peak $\phi_k$ and elevation peak $\theta_k$ are determined for each $k^{th}$ target.

\section{Performance Analysis}
\label{sec:Results}

\begin{figure*}[htbp]
    \centering
    \includegraphics[scale = 0.056]{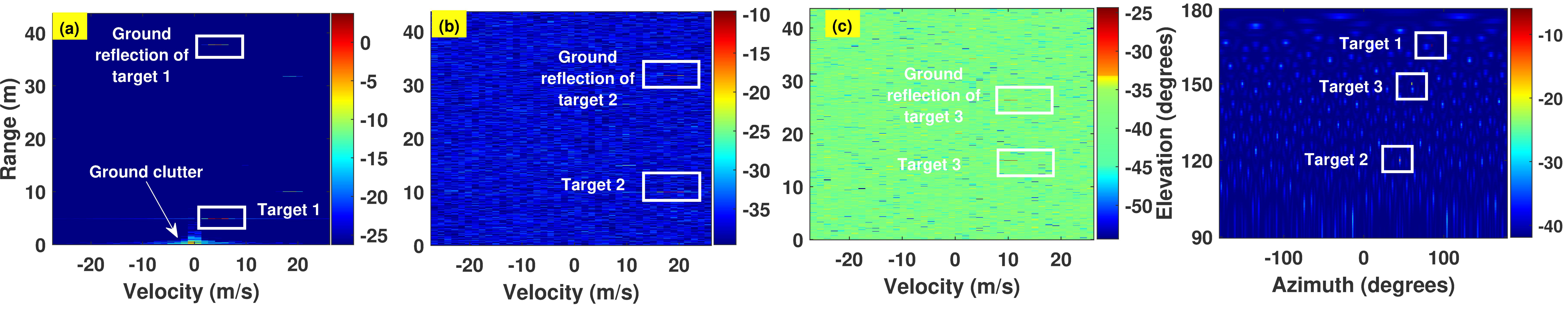}
    \includegraphics[scale = 0.056]{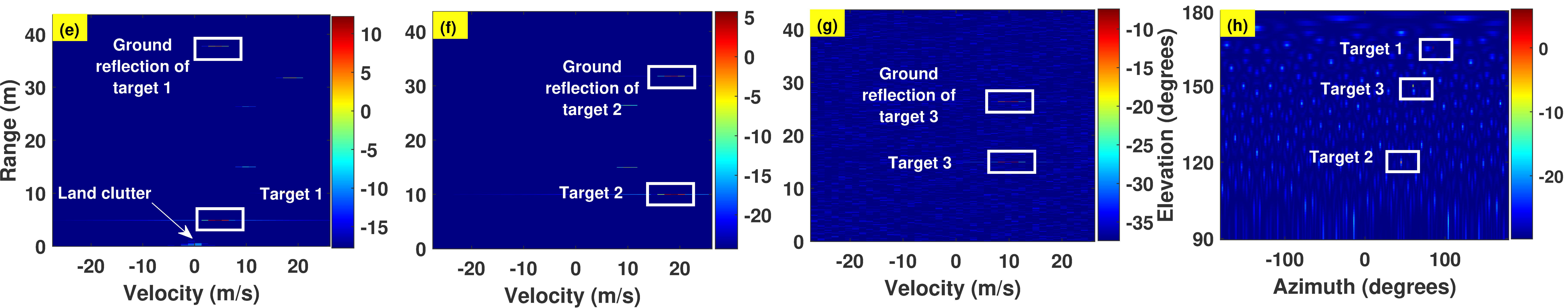}
    \caption{ 2D range-Doppler ambiguity map (a,e) without CLEAN, (b,f) after first CLEAN iteration, and (c,g) after second CLEAN iteration for horizontal and vertical polarization, respectively. 2D azimuth-elevation map for (d) horizontal and (h) vertical polarization.}
    \label{fig:rd_ae_all}
\end{figure*}

 We consider a three-dimensional spherical coordinate space to simulate the environment. Next, we consider an aerial base station transceiver mounted on a swarm of 8 UAVs corresponding to the antenna elements arranged in a UCA with the origin located at the center of the UCA. The radii of the UAVs from the origin is 1.07m. The UAVs are located at an elevation angle of 90$^o$ with respect to the z-axis, and the UAVs are distributed from 0$^o$ to 360$^o$ with an interval of 45$^o$. In this simulation experiment, we consider three targets, each modeled as an isotropic point scatterer. The RCS along vertical polarization, $\sigma_{VV}$, of the first, second, and third targets are set to 10, 5, and 1 square meters, respectively On the other hand, RCS along horizontal polarization $\sigma_{HH}$ of first, second and third targets are 2, 1 and 0.2 square meters, respectively. The first target is located at spherical coordinates [5m, 165$^o$, 95$^o$] and is moving with a velocity of 4m/s towards the base station. Further, the second and third targets are located at [10m, 120$^o$, 50$^o$] and [15m, 150$^o$, 70$^o$], respectively, and are moving towards the base station with a velocity of 18m/s and 10m/s, respectively. Next, we model the static ground clutter in a similar manner as shown in \cite{ram2022uav} with a clutter coefficient of -5dB.
 The rest of the parameters considered for the simulation setup are specified in Table~\ref{tab:RadarParam_Sim}. Note that the azimuth search spans from -180$^o$ to 180$^o$ and the elevation search space varies from 90$^o$ to 180$^o$.
\begin{table}[t]
    \centering
    \caption{ Simulated Radar Parameters}
    \begin{tabular}{p{5cm}|p{1cm}}
    \hline 
     \textbf{Parameter} & \textbf{Value}  \\
    \hline 
        Centre frequency ($f_o$) & 60GHz  \\ 
        Bandwidth ($BW$) & 1.76GHz  \\
        Pulse repetition interval ($T_{P}$) & 2$\mu$s \\
        Coherent processing interval ($T_{CPI}$) & 4ms \\
         Maximum unambiguous velocity  & 625 m/s\\
        Maximum unambiguous range  & 44m\\
        Velocity resolution & 0.3m/s\\
         Range resolution & 0.085m \\    
    \hline 
    \end{tabular}
    \label{tab:RadarParam_Sim}
\end{table}

Further, we discuss the simulation results for the localization of multiple targets in terms of range, Doppler velocity, azimuth, elevation, and sense of polarization. Firstly, we present the results for horizontal polarization as shown in Fig.~\ref{fig:rd_ae_all}(a-d). We observe the highest strength target being localized in the 2D range-Doppler ambiguity map in Fig.~\ref{fig:rd_ae_all}(a). The ground reflection of the target is also observed in the same Doppler bin and range bin corresponding to the ground reflected path. Note that the strength of the ground reflection is lower than the direct path due to higher path loss. Further, the static ground clutter is also present at the zero Doppler bin. In Fig.~\ref{fig:rd_ae_all}(b), we present the 2D range-Doppler ambiguity map after the first CLEAN iteration and static clutter filtering. Here, we observe that the second highest strength target is localized along with the ground reflection. Similarly, the third target, which has the lowest strength, is localized in terms of range and Doppler, as shown in Fig.~\ref{fig:rd_ae_all}(c) after the second CLEAN iteration. Further, we present the 2D azimuth-elevation map with all three targets localized in terms of azimuth and elevation as shown in Fig.~\ref{fig:rd_ae_all}(d). We observe here that the strength of the peaks of the indices indicating the targets are in accordance with the strength of the targets. 

Next, we present the results for vertical polarization in Fig.~\ref{fig:rd_ae_all}(e-h). Here, we observe that the strength of the peaks detected in the range-Doppler ambiguity map, as well as the 2D azimuth-elevation map, are much higher than those in the results of horizontal polarization as the targets dominantly scatter the V-V polarization. Note that the lowest strength of the target detected is approximately -25dB in horizontal polarization, resulting in lower SNR. On the other hand, the strength of the same target in vertical polarization is -8dB, which results in a considerably high SNR. Thus, the sense of polarization is helpful in detecting the lower strength targets.
 
\section{Summary}
\label{sec:Conclusion}
We demonstrated the 5-D sensing of mobile targets using an ISAC-enabled UAV swarm system in an urban environment. The simulation experiment conducted in this research validates the efficacy of the proposed system by accurately localizing multiple mobile targets based on range, Doppler velocity, azimuth, elevation, and polarization in the presence of ground clutter. This work lays the foundation for further research, including the extension of the system to more complex extended targets, such as pedestrians and vehicles. Additionally, future investigations could focus on analyzing the impact of the UAV swarm on communication link performance metrics, particularly in urban environments characterized by dynamic and multipath-rich conditions. Furthermore, the optimization of swarm coordination strategies and their influence on both sensing accuracy and communication efficiency warrant exploration to enhance the overall system's robustness and adaptability in real-world scenarios.

\bibliographystyle{IEEEtran}
\bibliography{references}

\end{document}